\documentclass[12pt]{article}
\usepackage{graphicx}
\usepackage[justification=justified, singlelinecheck=off, compatibility=false]{caption}
\usepackage{subcaption}
\usepackage{amsmath}
\usepackage{bm}
\usepackage{amssymb}
\usepackage{color}
\usepackage{float}
\usepackage[utf8]{inputenc}

\newcommand\pubnumber{CIPANP2018-Hyde}
\newcommand\pubdate{\today}

\def\address{Department of Physics \& Astronomy\\
Goucher College, Baltimore, Maryland 21204, USA}

\def\Title#1{\begin{center} {\Large #1 } \end{center}}
\def\Author#1{\begin{center}{ \sc #1} \end{center}}
\def\Address#1{\begin{center}{ \it #1} \end{center}}

\newcommand\pubblock{\rightline{\begin{tabular}{l} \pubnumber\\
         \pubdate  \end{tabular}}}
\newenvironment{Abstract}{\begin{quotation}  }{\end{quotation}}
\newenvironment{Presented}{\begin{quotation} \begin{center} 
             PRESENTED AT\end{center}\bigskip 
      \begin{center}\begin{large}}{\end{large}\end{center} \end{quotation}}
\def\Acknowledgements{\bigskip  \bigskip \begin{center} \begin{large}
             \bf ACKNOWLEDGEMENTS \end{large}\end{center}}

\def\beq{\begin{equation}}
\def\eeq#1{\label{#1}\end{equation}}
\def\eeqn{\end{equation}}

\def\beqa{\begin{eqnarray}}
\def\eeqa#1{\label{#1}\end{eqnarray}}
\def\eeqan{\end{eqnarray}}

\let\bar=\overbar

\def\Dslash{\not{\hbox{\kern-4pt $D$}}}
\def\dslash{\not{\hbox{\kern-2pt $\del$}}}

\def\msb{{\bar{\ssstyle M \kern -1pt S}}}

\newcommand{\fref}[1]{Fig.~\ref{#1}}
\newcommand{\sref}[1]{Sec.~\ref{#1}}

\begin{document}
\begin{titlepage}
\pubblock

\vfill
\Title{Detecting CP Violation in the Presence of Non-Standard Neutrino Interactions}
\vfill
\Author{Jeffrey M. Hyde}
\Address{\address}
\vfill
\begin{Abstract}
New physics beyond the Standard Model could appear at long baseline oscillation experiments as non-standard interactions (NSI) between neutrinos and matter. If so, determination of the CP-violating phase $\delta_{13}$ is ambiguous due to interference with additional complex phases. I'll present my work using both numerical solutions and a perturbative approach to study oscillation probabilities in the presence of NSI. I'll show how the CP phase degeneracies are visualized on biprobability plots, and the extent to which the energy spectrum for a given baseline length can help resolve them. In particular, this shows how the broad range of energies at DUNE would help distinguish between maximal, standard CP violation and the absence of CP violation with large $\epsilon_{e\tau}$.
\end{Abstract}
\vfill
\begin{Presented}
Thirteenth Conference on the Intersections of Particle and Nuclear Physics\\
Palm Springs, California, USA, May 29 -- June 3, 2018
\end{Presented}
\vfill
\end{titlepage}
\def\thefootnote{\fnsymbol{footnote}}
\setcounter{footnote}{0}

\section{Introduction}

The determination of CP violation in the leptonic sector is a target of current and upcoming neutrino experiments, and the result could have implications for early universe processes such as leptogenesis \cite{Branco:2011zb,Hagedorn:2017wjy}. Neutrino interactions violate CP invariance if oscillation probabilities in vacuum differ between neutrinos and antineutrinos, i.e. if $P(\nu_{\alpha} \rightarrow \nu_{\beta}) \neq P(\overline{\nu}_{\alpha} \rightarrow \overline{\nu}_{\beta})$ for flavors $\alpha$, $\beta$ with $\alpha \neq \beta$. In the standard 3-neutrino mixing scenario, the phase $\delta_{13}$ in the mixing matrix $U$ will give rise to CP violation unless its value is 0 or $\pi$.\footnote{Further background and terminology may be found in the paper this talk is based on \cite{Hyde:2018tqt}.} In this talk, I'll focus on a particular channel relevant to long-baseline experiments: $P(\nu_{\mu} \rightarrow \nu_{e}) \equiv P$ and $P(\overline{\nu}_{\mu} \rightarrow \overline{\nu}_{e}) \equiv \overline{P}$.

%
\begin{figure}[htb]
\centering
\includegraphics[width=0.62\textwidth]{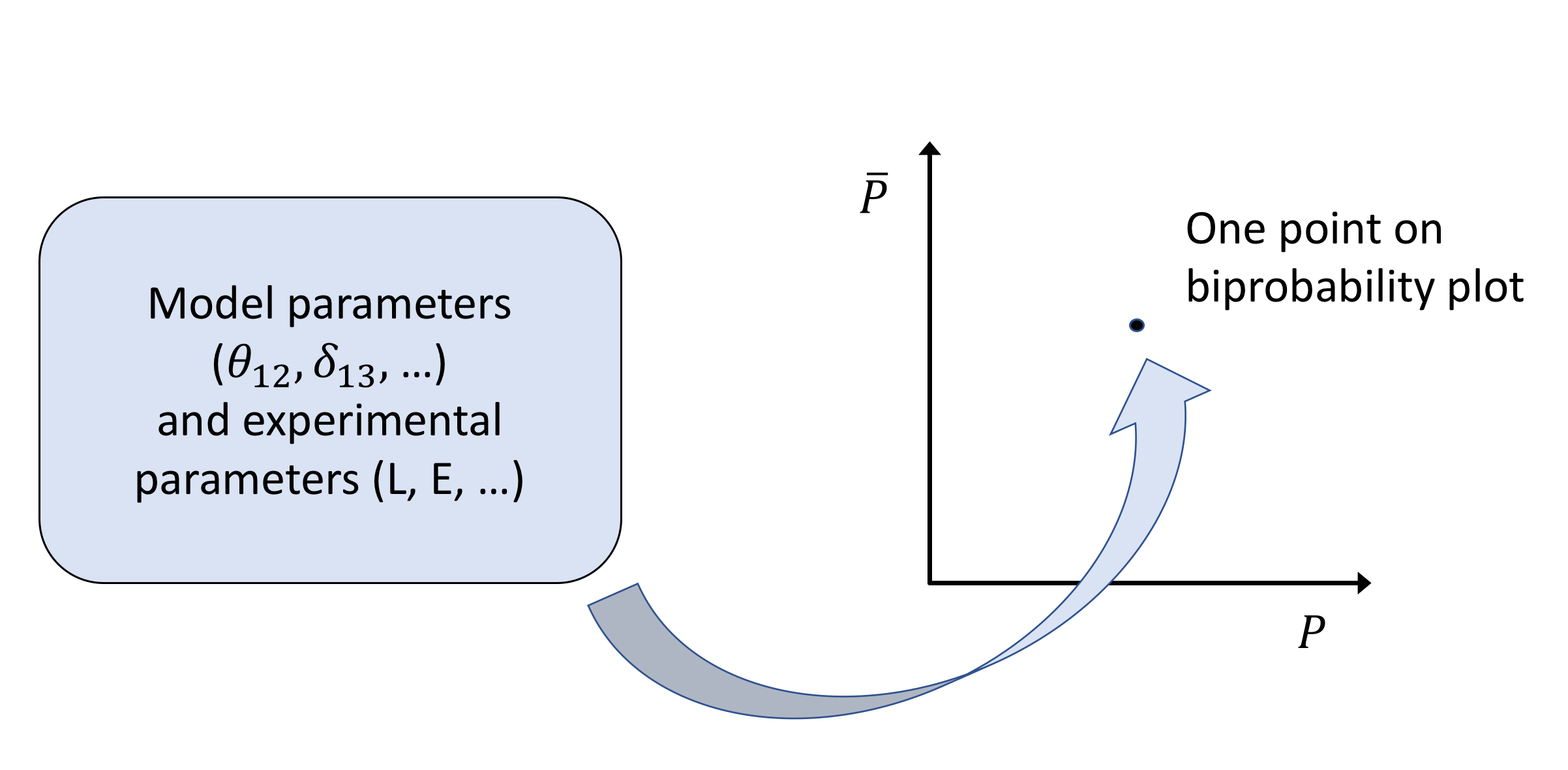} \, \, \, \includegraphics[width=0.32\textwidth]{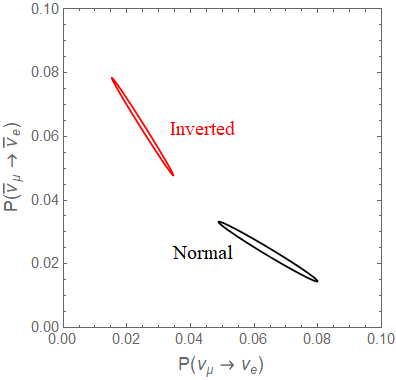}
\caption{\label{fig:biprob_point}A set of oscillation parameters predicts one point on the biprobability plot. Varying $\delta_{13}$ continuously from 0 to $2\pi$ traces out an ellipse in biprobability space. In vacuum, the normal and inverted hierarchy ellipses nearly overlap, but propagation through matter breaks this degeneracy.}
\end{figure}

The above description makes neutrino CP violation seem straightforward: in vacuum, if $P \neq \overline{P}$, then $\delta_{13} \neq 0, \pi$. But other effects can complicate this:
\begin{itemize}
 \item Long-baseline experiments: Neutrino beams pass through Earth, so the potential is not CP-invariant.
 \item Non-standard neutrino interactions (NSI): Off-diagonal $\epsilon_{\alpha\beta}$ in the matter potential will contain a new CP-violating phase \cite{Grossman:1995wx,Friedland:2004pp,Friedland:2004ah,Antusch:2008tz,Miranda:2015dra,Farzan:2015hkd}.
\end{itemize}
Much previous work has examined degeneracies in neutrino oscillations with and without NSI.\footnote{See \cite{Arafune:1997hd,Barger:2001yr,Minakata:2001qm,GonzalezGarcia:2001mp,Coloma:2011rq,Friedland:2012tq,Rahman:2015vqa,Masud:2015xva,Coloma:2015kiu,Palazzo:2015gja,deGouvea:2015ndi,Masud:2016bvp,Liao:2016hsa,deGouvea:2016pom,Liao:2016orc,Ge:2016dlx,Agarwalla:2016fkh,Fukasawa:2016lew,Deepthi:2016erc,Forero:2016cmb,deGouvea:2017yvn,Deepthi:2017gxg,Flores:2018kwk}.} The question I'm concerned with here is: in the presence of NSI, how can we tell whether CP is violated and determine the underlying parameters? This talk is based on \cite{Hyde:2018tqt}, where further details of calculations and notation may be found. I'll focus on nonzero $\epsilon_{e\mu}$ or $\epsilon_{e\tau}$, which have very similar effects as far as this work is concerned, although the constraints are looser on $\epsilon_{e\tau}$. (See e.g. \cite{Ohlsson:2012kf}.)

\section{Degeneracies and Their Resolution}\label{sec:degeneracy}

\subsection{Degeneracies on Biprobability Plot}

For a given set of parameters ($\theta_{12}$, $\delta_{13}$, etc. as well as experimental $L$, $E$, etc.) the value of $P$ and $\overline{P}$ are determined. This leads to one point on the biprobability plot; see Fig.~\ref{fig:biprob_point}. Varying $\delta_{13}$ from 0 to $2\pi$ produces a closed curve that happens to be elliptical, and was originally used to show how matter effects lift an approximate degeneracy between the normal and inverted mass orderings \cite{Minakata:2001qm}.

%
\begin{figure}[htb]
\centering
\includegraphics[width=0.4\textwidth]{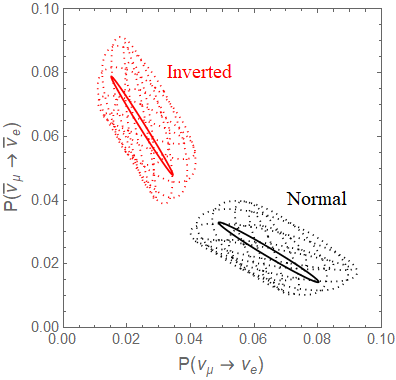}
\caption{\label{fig:L1300_ellipse_fuzzball}In the presence of non-standard neutrino interactions (NSI), degeneracies get worse. Solid curves show the standard case, dotted curves show some of the possibilities if $|\epsilon_{e\mu}| = 0.05$.}
\end{figure}

In the presence of NSI, there is an increased degree of degeneracy, as shown in \fref{fig:L1300_ellipse_fuzzball}. This looks like a mess, but the perturbative representation of oscillation probabilities (which treats $\Delta m^2_{21} / \Delta m^2_{31}$, $\theta_{13}$, $|\epsilon_{\alpha\beta}|$ as small parameters in computing $P$, $\overline{P}$) lends insight that may be shown more precisely with numerical results \cite{Arafune:1997hd,Cervera:2000kp,Kikuchi:2008vq,Asano:2011nj}. One could use this to quantitatively describe how the ellipses are streteched and rotated as seen in \fref{fig:L1300_ellipse_fuzzball}. However, the more practical question is ``given a measured point (or region) in biprobability space, is there an easy way to tell which degeneracies are excluded or not?"

One feature of the perturbative expressions is that the neutrino oscillation probability can be written schematically as $P = \, $(CP-even terms) + (CP-odd terms), so the antineutrino oscillation probability is $\overline{P} = \, $(CP-even terms) $-$ (CP-odd terms). Therefore it is natural that the rotated coordinates $P^+ \equiv \frac{1}{\sqrt{2}}(P + \overline{P}) \sim $ (CP-even terms) and $P^- \equiv \frac{1}{\sqrt{2}}(P - \overline{P}) \sim $ (CP-odd terms) will be useful.\footnote{Note that CP-even terms can include pairs of CP-violating phases, etc.}

\begin{figure}[htb]
\centering
\includegraphics[width=0.4\textwidth]{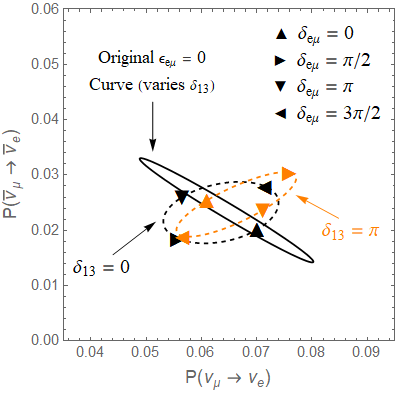} \, \, \, \, \, \includegraphics[width=0.4\textwidth]{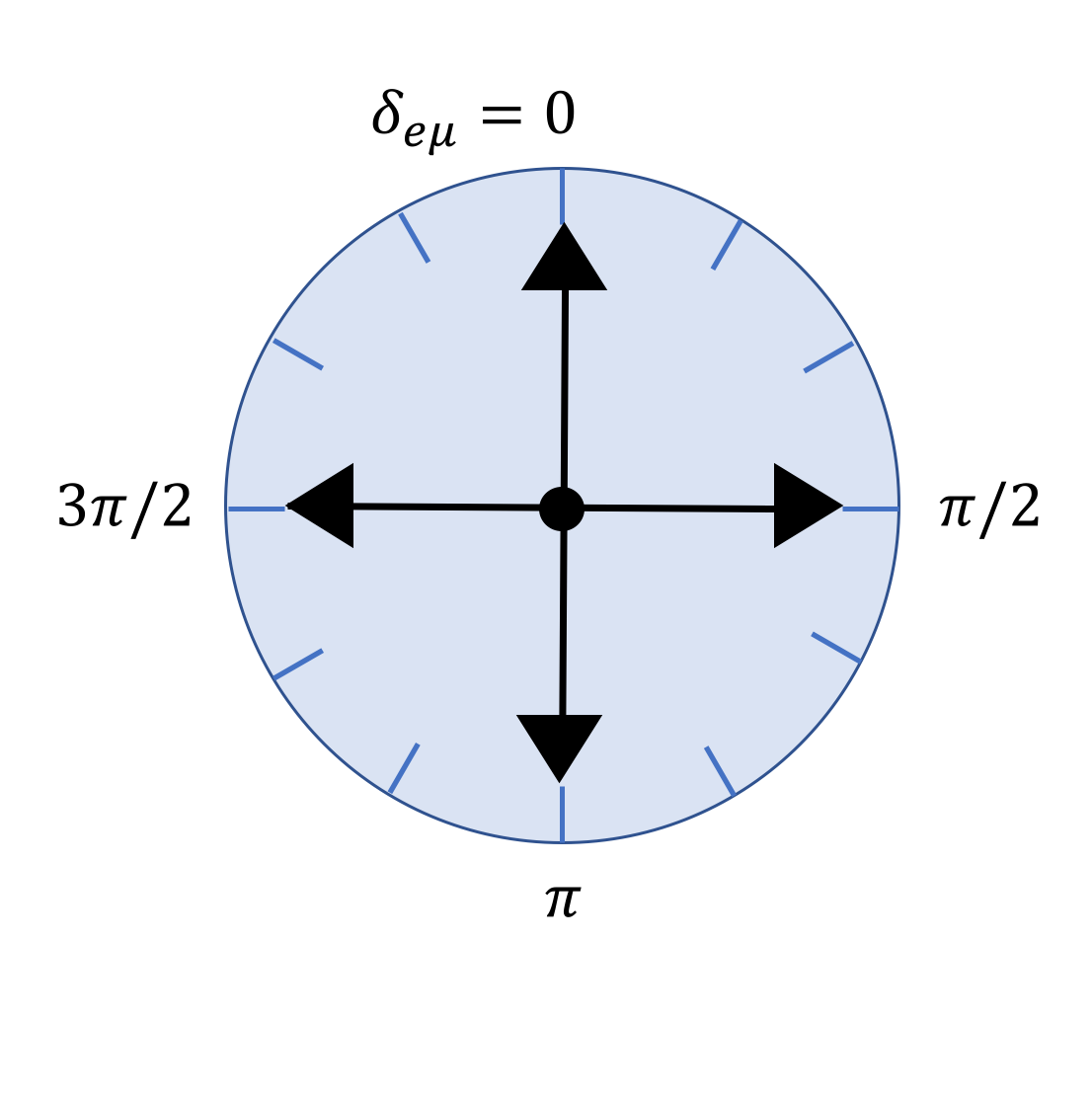}
\caption{Two ``hidden sector" ellipses (varying $\delta_{e\mu}$) that are approximately degenerate, superposed over the original biprobability curve that varies $\delta_{13}$. The values of $\delta_{e\mu}$ are shown by the triangles, which can be thought of as ``hands on a clock" that go around once per $2\pi$.}
\label{fig:nsi_degen_example}
\end{figure}

An interesting consequence of the perturbative expansion at $\Delta_{31} = \pi/2$\footnote{$\Delta_{31} \equiv \Delta m^2_{31} L / (4 E ) \sim \pm 0.003 \, L$[km] / $E$[GeV]} is that dominant terms give
\begin{align}
P^+ &\approx 0.060, \, \, \, \, \, P^- \, \approx \, 0.027 - 0.017\sin(\delta_{13})
\end{align}
in the absence of NSI.\footnote{The specific numbers depend on the experimental parameters (numbers given relevant to DUNE) but qualitative behavior does not. Full expressions in \cite{Hyde:2018tqt}.} Points with a given $\delta_{13}$ are approximately degenerate with points that correspond to $\pi - \delta_{13}$. As a consequence, curves drawn with $\delta_{e\mu}$ varying from 0 to $2\pi$ (now $\delta_{13}$ fixed) -- also elliptical -- roughly overlap for identical values $\delta_{13}$ and $\pi - \delta_{13}$. For this reason, the ``hidden sector" ellipse turns out to be most useful for representing degeneracies. (See also \cite{Gago:2009ij,Soumya:2016enw,Agarwalla:2016fkh,Forero:2016cmb}.) At leading order, these hidden sector ellipses have a center determined by $\sin(\delta_{13})$.

The effect of nonzero $\epsilon_{e\mu}$ or $\epsilon_{e\tau}$ and $\delta_{e\mu}$ or $\delta_{e\tau}$ on a point corresponding to some $\delta_{13}$ is to move it in the $P^+$ and/or $P^-$ direction in a way that depends mainly on $\delta_+ \equiv \delta_{13} + \delta_{e\tau}$:
\begin{align}
\Delta P^+_{e\mu, e\tau} \approx 0.21 |\epsilon_{e\mu, e\tau}| \sin(\delta_+), \, \, \, \, \, \Delta P^-_{e\mu, e\tau} \, \approx \, 0.14 |\epsilon_{e\mu}|\sin(\delta_+), \, \, 0.13 |\epsilon_{e\tau}|\sin(\delta_+).
\end{align}
As $\delta_+$ varies from 0 to $2\pi$, the range of $P^{\pm}$ values vary between $\Delta P^+ \sim \pm 0.21 |\epsilon_{e\mu, e\tau}|$ and $\Delta P^- \sim \pm (0.14, \, 0.13) |\epsilon_{e\mu, e\tau}|$. These can be thought of as an approximate ``width" of the hidden sector ellipse in the $P^{\pm}$ directions.

\fref{fig:nsi_degen_example} shows an example of overlapping hidden sector ellipses, each varying $\delta_{e\mu}$ with fixed $\delta_{13} = 0$ or $\pi$ for $|\epsilon_{e\mu}| = 0.05$. The triangles represent different values of $\delta_{e\mu}$ while each hidden sector ellipse has fixed $\delta_{13}$. This is a lot of information to represent on one plot, but for convenience the orientation of the triangles can be imagined as ``hands on a clock" that go around once per $2\pi$. The original ellipse that varies $\delta_{13}$ with zero NSI is also shown for comparison.


\subsection{Degeneracy Breaking}

\begin{figure}[t]
\centering
\includegraphics[width=0.4\textwidth]{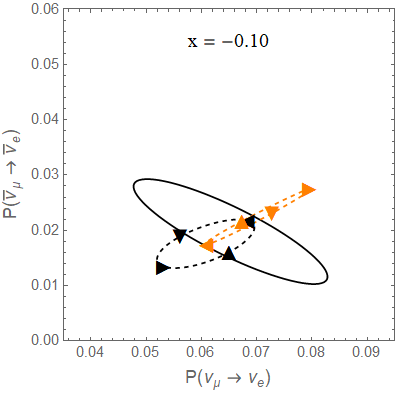} \, \, \, \, \, \includegraphics[width=0.4\textwidth]{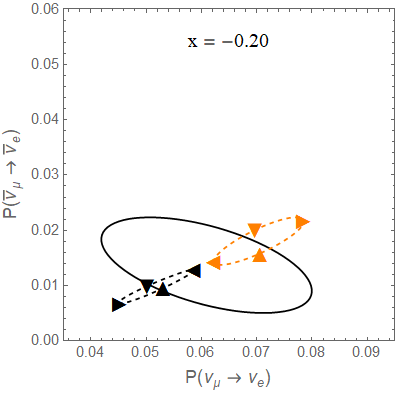}
\caption{\label{fig:ellipse_variation_offpeak_nh}Varying the neutrino energy by decreasing from $\Delta_{31} = \pi/2$ separates overlapping hidden sector ellipses along the $P^+$ direction. (Compare with \fref{fig:nsi_degen_example}, which shows the $\Delta_{31} = \pi/2$ $(x = 0)$ situation.)}
\end{figure}

Degeneracies can be broken by a different baseline length and/or neutrino energy. In the perturbative expansion, the $\mu \rightarrow e$ oscillation probability is near a maximum for $\Delta_{31} \equiv \Delta m^2_{31} L / 4E = \pi/2$, which depends on the ratio $L/E$. There is also a matter effect piece that depends linearly on $L$. Some experiments (such as T2K) have a narrow energy band while others (such as DUNE) will have a wider range of neutrino energies. For a broad spectrum, it's convenient to expand around the energy $E_0$ for which $\Delta_{31} = \pi/2$ for a given experiment, and parametrize the neutrino energy as $E = E_0 \left( 1 + x \right)$, i.e. $x$ is the fractional change in energy from $E_0$. On a biprobability plot, decreasing the energy from $E_0$ ($x<0$) tends to separate degenerate points in the $P^+$ direction, improving the possibility for resolving different parameters. For instance,
\begin{align}
	P^+ &\approx 0.059 + 0.21|\epsilon_{e\tau}|\sin(\delta_+) + 0.027 x \cos(\delta_{13})
\end{align}
for $L = 1300$ km and $\rho = 3$ g/cm$^3$. For $x = 0$ only $\delta_+ = \delta_{13} + \delta_{e\tau}$ controls the degeneracy in the $P^+$ direction, but for $x \neq 0$ the $\cos(\delta_{13})$ term comes into play. Returning to the example of $\delta_{13} = 0$ versus $\pi$, the cosine will equal $+1$ and $-1$ respectively, moving these points in opposite directions along $P^+$.

\section{Degeneracy Breaking for Apparent $\delta_{13} = -\pi/2$}\label{sec:applications}

One application of \sref{sec:degeneracy} is the situation where $P$, $\overline{P}$ would be measured as apparently consistent with $\delta_{13} = -\pi/2$. This is experimentally motivated by T2K and NO$\nu$A results that suggest $\delta_{13} \sim -\pi/2$, although no values have yet been ruled out \cite{Abe:2017vif,NOvA:2018gge}.

The larger $\epsilon_{e\tau}$ is, the more $\delta_{13}$ can vary away from the apparent value of $-\pi/2$. In particular, apparent $\delta_{13} = -\pi/2$ is also consistent with $|\epsilon_{e\tau}| = 0.2$ and $\delta_{13}, \delta_{e\tau} = 0$ or $\delta_{13}, \delta_{e\tau} = \pi$. For experimental parameters of DUNE, these points are more easily distinguishable than at NO$\nu$A. In particular, both numerical results and the perturbative approach in this paper show that the $\delta_{13}, \delta_{e\tau} = \pi$ point becomes well-separated as energy varies, but the $\delta_{13}, \delta_{e\tau} = 0$ point remains more difficult to distinguish from maximal CP violation in the standard scenario.

\begin{figure}[t]
\centering
\includegraphics[width=0.3\textwidth]{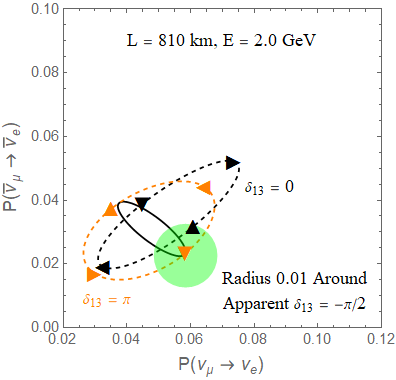} \, \, \, \includegraphics[width=0.3\textwidth]{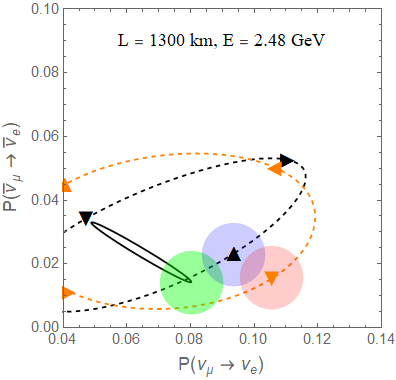} \, \, \, \includegraphics[width=0.3\textwidth]{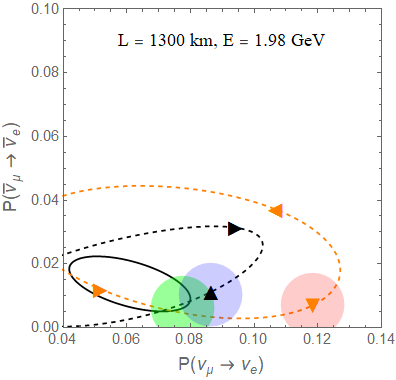}
\caption{\label{fig:maxcpv_nova_dune}Degeneracy for apparent $\delta_{13} = -\pi/2$ at NO$\nu$A is somewhat broken by DUNE (for $\Delta_{31} = \pi/2$, i.e. $x=0$) and further by DUNE with $x = -0.2$ ($E  = 1.98$ GeV).}
\end{figure}

\fref{fig:maxcpv_nova_dune} shows a few points with ``uncertainty disks" of biprobability radius 0.01 drawn around them.\footnote{This is not a precise statement about what each experiment can resolve, but a convenient way to represent the approximate order at which points may or may not be easily resolved.} For apparent $\delta_{13} = -\pi/2$ at NO$\nu$A, the point for $|\epsilon_{e\tau}| = 0.2$, $\delta_{13} = \pi$, $\delta_{e\tau} = \pi$ lies near the center of this disk, and $|\epsilon_{e\tau}| = 0.2$, $\delta_{13} = 0$, $\delta_{e\tau} = 0$ lies away from the center but within the disk. To within $\Delta P^{\pm} \approx 0.01$, the three points are indistinguishable. At DUNE (with $x=0$), these points separate in the $P$ direction. However, there is still some overlap in the uncertainty disk around each point. With $x = -0.2$, however, the point corresponding to $\delta_{13}, \delta_{e\tau} = \pi$ separates from the other two along the $P$ direction. This pattern of degeneracy breaking is described by the perturbative expressions discussed in \cite{Hyde:2018tqt}.

Together, the numerical results and the perturbation expansion illustrate the role of a broad energy spectrum in addition to different baseline lengths in breaking specific degeneracies. This provides a useful way to characterize the role of each parameter in determining which degeneracies are present, and how those may be lifted.

\Acknowledgements
I am grateful to the organizers of CIPANP 2018 for the opportunity to speak.

\end{document}